# *Boletín*
## de la
## Sociedad Mexicana de Física

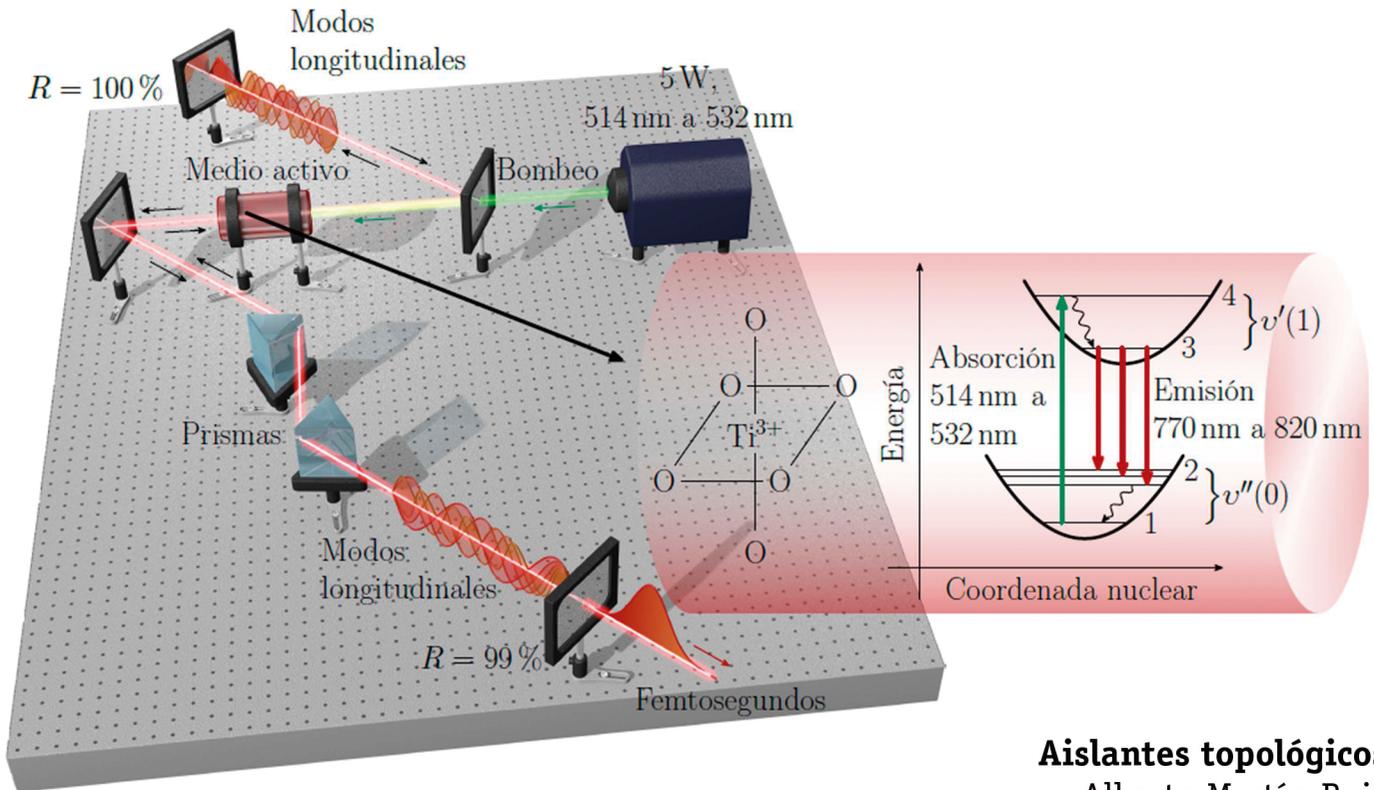

**Aislantes topológicos**
Alberto Martín-Ruiz
Instituto de Ciencias Nucleares, UNAM

**El Sistema Internacional de Unidades a 230 años de evolución**
Eric Rosas[a] y Niels Rosas
Clúster Mexicano de Fotónica, A. C.
[a]Vicepresidente en la International Commission for Optics.

**El nuevo sistema láser de pulsos ultracortos de la UAM-Iztapalapa**
Luis Guillermo Mendoza Luna[a,b], César A. Guarín-Durán[a,b]
Emmanuel Haro Poniatowski[a], José Luis Hernández Pozos[a]
[a]Departamento de Física, UAM-Iztapalapa
[b]Catedrático CONACYT

distinciones ▫ reseñas ▫ efemérides ▫ placeres del pensamiento ▫ noticias de la comunidad ▫ bibliografías

# Aislantes topológicos

### *Alberto Martín-Ruiz*
Instituto de Ciencias Nucleares, Universidad Nacional Autónoma de México

**Introducción**. La física de materia condensada estudia sistemas con un gran número de átomos interactuantes que dan lugar a estados agregados. El concepto de emergencia, introducido por Philip W. Anderson y desarrollado por Robert Laughlin, es la piedra angular sobre la que descansa la descripción de estos estados colectivos de la materia. Éste se refiere a que las propiedades de un sistema compuesto no se pueden inferir a partir del conocimiento de las propiedades de sus componentes, sino que emergen como consecuencia de la organización espontánea de éstos. Como ejemplo representativo de una propiedad emergente de la materia condensada podemos mencionar la superconductividad, en donde los electrones de conducción dejan de comportarse como electrones independientes y forman un estado cuántico colectivo macroscópico. Otro ejemplo de propiedad emergente son las *cuasipartículas*, que surgen como consecuencia del comportamiento colectivo de los electrones en ciertos materiales como el grafeno.

La mayoría de los estados o fases de la materia condensada se han logrado entender gracias al concepto de *rompimiento espontáneo de simetría*, introducido por Lev D. Landau a mediados del siglo pasado. Éste se puede ilustrar con los siguientes ejemplos. Un sólido cristalino rompe la simetría de traslación, a pesar de que la interacción entre sus componentes atómicos es traslacionalmente invariante. Un material ferromagnético rompe la simetría rotacional, aún cuando sus interacciones fundamentales son isotrópicas. El patrón de rompimiento espontáneo de simetría define un único parámetro de orden, que toma un valor esperado no nulo sólo en el estado ordenado, y es posible formular una teoría de campos efectiva (llamada teoría de Landau-Ginzburg) basada en propiedades generales tales como la dimensionalidad y la simetría de dicho parámetro de orden. Por ejemplo, el parámetro de orden relevante en la fase magnética de un material ferromagnético corresponde a la magnetización local, es decir, la distribución espacial del promedio de espín de los electrones que componen el material. A altas temperaturas, en la fase no magnética, este parámetro de orden es cero, pero se vuelve no nulo tan pronto el sistema se ordena magnéticamente al enfriarse por debajo de la temperatura de Curie.

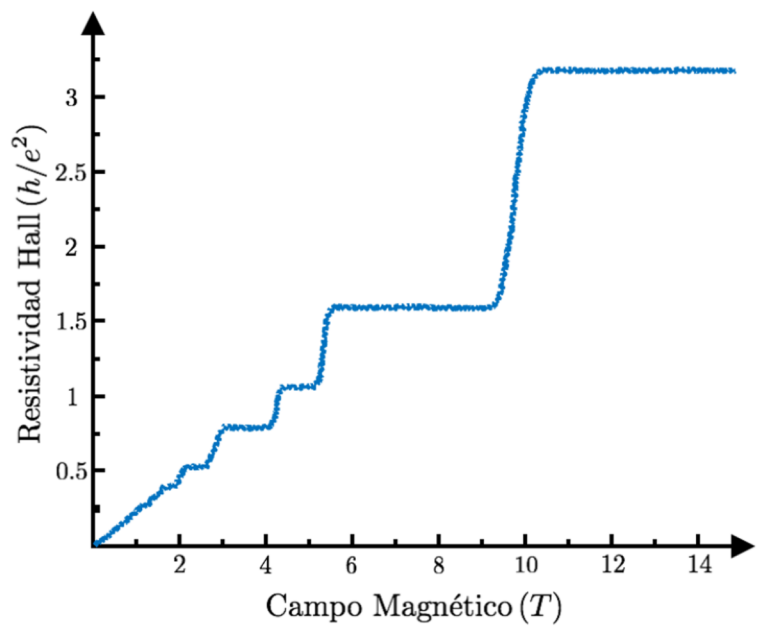

*Figura 1: Resistividad Hall en función del campo magnético.*

**Efecto Hall cuántico**

Parecía que una de la grandes metas de la física de materia condensada, la de entender y clasificar las distintas fases de la materia, había sido alcanzada. Sin embargo, Klaus von Klitzing descubrió en 1980 un nuevo estado cuántico, el *estado Hall cuántico* (HC), que elude el esquema de clasificación de Landau. En su experimento, von Klitzing sometió un gas de electrones a muy baja temperatura, confinado en un plano a un fuerte campo magnético perpendicular a la muestra. Lo que observó fue que la conductividad Hall estaba cuantizada en múltiplos enteros de la unidad fundamental de conductividad, esto es

$$Hall \quad n(e^2/h). \tag{1}$$

En la Fig. 1 se muestra la resistividad Hall, tal como la midió von Klitzing. En el estado HC, la superficie de la muestra bidimensional es aislante, y la corriente eléctrica se transporta sin disipación sólo en el borde. La explicación de por qué la cuantización de la conductividad Hall es tan precisa (¡una parte en un billón!) e insensible a pequeños cambios en la muestra, vino años más tarde y tiene que ver con la emergencia de una nueva fase de la materia que cae fuera del paradigma de Landau, pues ésta no surge a partir del rompimiento espontáneo de simetrías. De esta forma, el estado HC fue el primer ejemplo de una fase de la materia que es topológicamente distinta de los otros estados de la materia conocidos. Para entender las propiedades del estado HC y sus diferencias con el estado aislante convencional, repasemos brevemente sus características.

El estado aislante es el estado más simple de la materia. Como se sabe por la experiencia cotidiana, los aislantes son materiales que no conducen bien la electricidad. En un *aislante atómico*, los electrones están fuertemente ligados a los átomos en capas cerradas, de modo que costaría energía hacerlos conducir. A la energía mínima necesaria para promocionar los electrones de estos estados de valencia a los estados de conducción, se le denomina *brecha* de energía (o banda prohibida). Los aislantes tienen una brecha grande, típicamente mayor a 1eV, de modo que sus electrones requieren grandes cantidades de energía para moverse. Ver el panel superior de la Fig. 2. Aunque la brecha de energía en un aislante atómico (tal como el argón sólido) es más grande que la de un semiconductor, en cierto sentido ambos estados pertenecen a la misma fase de la materia. Esta equivalencia podemos pensarla como una transformación continua entre los Hamiltonianos de ambos estados sin cerrar la brecha de energía. De forma simple, podemos imaginar que encendemos (o apagamos) poco a poco el traslape entre los electrones de distintos átomos sin cerrar la brecha.

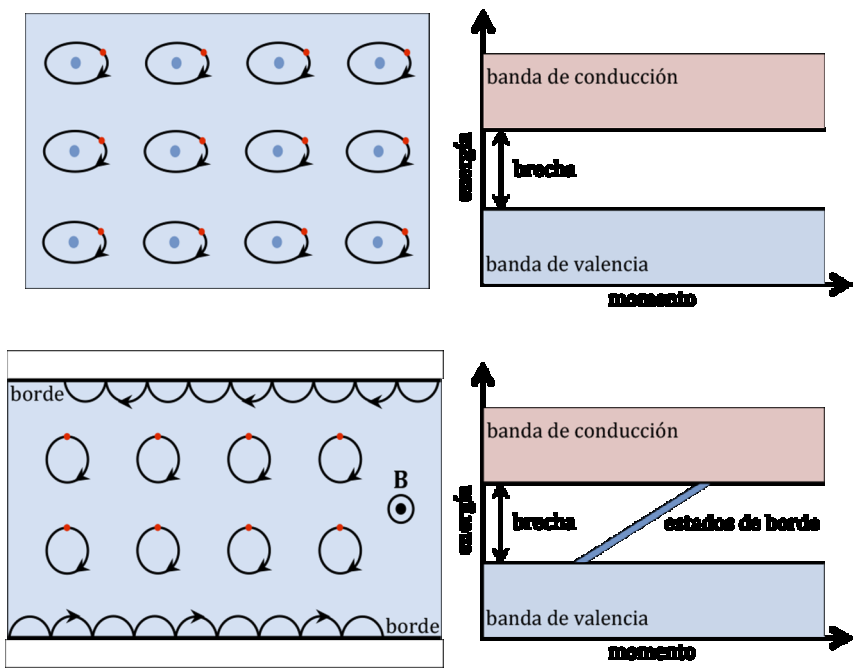

*Figura 2. Representación esquemática de un aislante convencional y su estructura de bandas (panel superior); y del efecto Hall cuántico y su estructura de bandas (panel inferior).*

El estado Hall cuántico podemos entenderlo de forma simple considerando el movimiento de electrones en una muestra bidimensional sujeta a un campo magnético fuerte. El campo hace que los electrones experimenten una fuerza de Lorentz que los hará moverse en órbitas circulares con frecuencia ciclotrón $\omega_c \sim \sqrt{eB/m}$, similar al movimiento periódico de los electrones atómicos. El tratamiento mecano-cuántico de este problema reemplaza el movimiento circular por orbitales con niveles de Landau cuantizados con energía $E_n \sim \hbar \omega_c (n + 1/2)$. Los niveles de Landau pueden interpretarse como una estructura de bandas: si $n$ niveles están llenos y el resto vacíos, entonces una brecha de energía separa a los estados ocupados y vacíos, justo como en un aislante. En las fronteras del sistema, sin embargo, los portadores de carga exhiben un tipo de movimiento distinto pues no son capaces de realizar órbitas cerradas, de manera que pueden rebotar y propagarse a lo largo del borde, como se muestra en el panel inferior de la Fig. 2. En la teoría cuántica, estas órbitas saltarinas corresponden a estados electrónicos quirales en el sentido de que se propagan unidireccionalmente en el borde y no tienen energías cuantizadas. Estos estados metálicos, diferentes de los demás estados ordinarios de la materia, hacen que el transporte electrónico sea perfecto: normalmente, los electrones son dispersados por impurezas (lo que da origen a la resistencia eléctrica), pero como no hay modos que se propaguen hacia atrás, los electrones no tienen otra elección más que propagarse hacia adelante. Esto lleva a lo que se conoce como transporte sin disipación de los estados de borde (no hay dispersión de electrones y por lo tanto no se pierde energía en forma de calor) y son los responsables de la cuantización precisa de la conductividad Hall. El mecanismo de transporte sin disipación podría ser extremadamente útil en dispositivos semiconductores. Sin embargo, el requerimiento de un campo magnético fuerte limita severamente las aplicaciones potenciales del efecto HC.

La pregunta crucial en esta historia es: ¿Cuál es la diferencia fundamental entre el estado Hall cuántico caracterizado por la Ec. (1) y un aislante ordinario? La respuesta es cuestión de topología, como lo explicaron Thouless, Kohmoto, Nightingale y den Nijs (TKNN) en 1982. El estado HC evade el esquema de clasificación de Landau porque no hay ningún rompimiento espontáneo de simetría. Sin embargo, existe una cantidad macroscópica medible análoga a los parámetros de orden, la conductividad Hall, que tiene ciertas propiedades muy peculiares: está cuantizada, no depende de la geometría, no se ve afectada por pequeñas variaciones del campo magnético y es in-

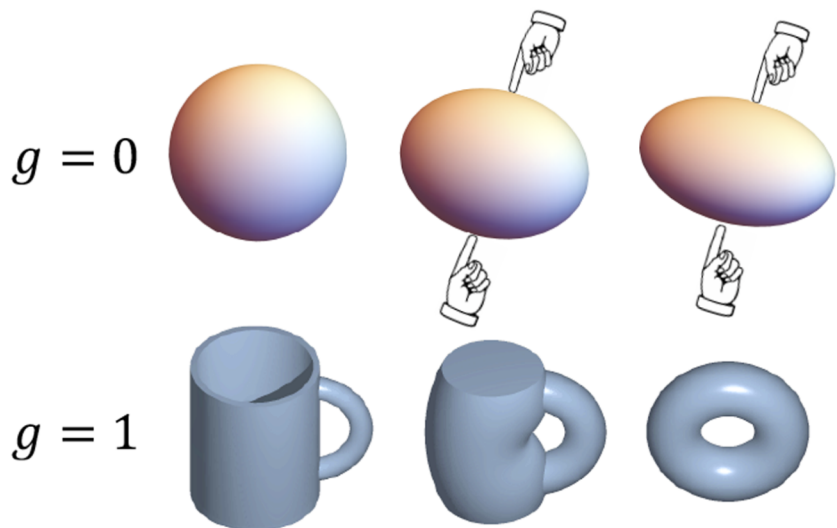

Figura 3: Equivalencia topológica entre una esfera y una elipse (panel superior); y entre una taza y una rosquilla (panel inferior).

# Artículos

sensible a impurezas en la muestra. Echemos un vistazo al origen topológico de la conductividad Hall.

**Invariantes topológicos**

La topología es la rama de las matemáticas que estudia qué propiedades de los objetos permanecen invariantes cuando los deformamos de manera suave. En este sentido, la topología se olvida de los detalles geométricos (locales) de un objeto, concentrándose sólo en sus propiedades globales. Los matemáticos introdujeron el concepto de *invariancia topológica* para clasificar los diferentes cuerpos geométricos en clases. Por ejemplo, las superficies bidimensionales se clasifican por un número entero $g$, el *invariante topológico*, que corresponde al número de *agujeros* o *genus* en ellas. Éste se calcula de manera rigurosa mediante la integral de superficie de la curvatura de Gauss ($k$):

$$\frac{1}{2\pi} \int_S k \, dS = 2 - 2g. \quad (2)$$

Mientras que el integrando depende de los detalles de la geometría de la superficie $S$, el teorema de Gauss-Bonnet garantiza que el valor de la integral es independiente de esos detalles y depende sólo del número de agujeros en ella. Objetos con el mismo valor de $g$ pertenecen a la misma clase topológica. Bajo esta clasificación, la superficie de una esfera perfecta es topológicamente equivalente a la superficie de un elipsoide, pues éstas pueden deformarse suavemente una en la otra sin crear ningún agujero. Del mismo modo, una taza de café es topológicamente equivalente a una rosquilla, pues ambas tienen el mismo número de agujeros. Ver la Fig. 3.

Los conceptos anteriores no se restringen al ámbito de la geometría. En física de materia condensada, la invariancia topológica no se relaciona con la forma geométrica del material, sino con el comportamiento de las funciones de onda de Bloch al recorrer la zona de Brillouin (que definimos como una región finita en el espacio de momentos del orden de $2\pi/a$ en cada dirección, donde $a$ denota la distancia típica entre dos nodos de la red). El concepto clave que proporciona la liga entre la física y la topología es el de *deformación suave*. En matemáticas, uno considera deformaciones suaves de formas geométricas sin acciones violentas capaces de crear agujeros. En física, podemos considerar un Hamiltoniano general de muchas partículas con una brecha de energía que separa el estado base de los estados excitados. En este caso, es posible definir una deformación suave como un cambio adiabático en el Hamiltoniano que no cierra la brecha de energía en el volumen. De manera similar a la clasificación topológica de las formas geométricas a través del genus, en física de materia condensada podemos clasificar la estructura de bandas de un material por medio del *número de Chern*, que es un invariante topológico definido en términos de las funciones de onda de Bloch en la zona de Brillouin. Este concepto topológico puede aplicarse tanto a aislantes como a superconductores con una brecha de energía completa, pero no es aplicable a estados sin brecha como los metálicos. De acuerdo a esta definición general, un estado con brecha no puede transformarse en otro que pertenezca a una fase topológica distinta, a menos que ocurra una transición de fase en donde el sistema se vuelva sin brecha. Lo anterior implica que un material topológico siempre tiene una frontera metálica (sin brecha) cuando está en contacto con un material trivial o el vacío.

Para entender mejor los conceptos anteriores, consideremos los estados electrónicos de un sólido clasificados en términos del momento cristalino $k$. Los estados de Bloch $|u_m(k)\rangle$, definidos en una sola celda unitaria del cristal, son estados propios del Hamiltoniano de Bloch $H(k)$. Los valores propios $E_m(k)$ definen las bandas de energía, que colectivamente forman la estructura de bandas del sólido. Es decir, una estructura de bandas bidimensional consiste en un mapeo del momento del cristal $k$ (definido sobre un toroide) al Hamiltoniano de Bloch $H(k)$. Estructuras de banda con brecha pueden clasi-ficarse topológicamente considerando las clases de equivalencia de $H(k)$ que pueden deformarse continuamente una en otra sin cerrar la brecha. Estas clases se distinguen por el invariante topológico $n \in \mathbb{Z}$ llamado número de Chern. Así, de la misma forma en que no es posible convertir una esfera (*con g = 0*) en una rosquilla (*con g = 1*) sin abrir un agujero, no podemos convertir una banda convencional ($n = 0$) en una topológica ($n \neq 0$) sin cerrar una brecha. Este concepto topológico es el que establece la diferencia entre el estado aislante convencional y el estado Hall cuántico.

El invariante de Chern tiene sus raíces en la teoría matemática de haces fibrados, pero puede entenderse físicamente en términos de la fase de Berry asociada con las funciones de onda de Bloch $|u_m(k)\rangle$. Siempre que no haya degeneración accidental, cuando $k$ se transporte a lo largo de un circuito cerrado en la zona de Brillouin, $|u_m(k)\rangle$ adquirirá una fase de Berry bien definida dada por la integral de línea de la conexión $A_m = i\langle u_m(k)|\nabla_k u_m(k)\rangle$. Ésta puede escribirse como una integral de superficie de la curvatura de Berry $F_m = \nabla_k \times A_m$. El invariante de Chern es el flujo total en la zona de Brillouin,

$$n_m = \frac{1}{2\pi} \int_{BZ} F_m \, d^2k, \qquad (3)$$

donde $n_m$ es un entero cuantizado por razones análogas a la cuantización de los monopolos magnéticos de Dirac. TKNN demostraron que cada banda ocupada contribuye con un entero $n_m$ a la conductividad Hall cuando se calcula usando la fórmula de Kubo. Es decir, ellos obtuvieron que la conductividad Hall es $n(e^2/h)$, donde $n = \sum_{m=0}^{N} n_m$ es el número de Chern total que se obtiene sumando los de cada banda ocupada. El número de Chern $n$ es un invariante topológico en el sentido de que no cambia cuando alguno de los parámetros del Hamiltoniano varía lentamente. Físicamente, el número de Chern determina el número de canales unidireccionales (quirales) que se propagan por el borde del material. Debido a su quiralidad, estos canales topológicos son robustos frente al desorden y conducen la corriente eléctrica sin pérdidas. Esto ayuda a explicar la cuantización robusta de $\sigma_{Hall}$.

**Aislantes topológicos**

El estado Hall cuántico pertenece a una clase topológica en donde la simetría de inversión temporal (IT) está rota, por ejemplo, por la presencia de un campo magnético. Sin embargo, en los últimos años hemos aprendido que pueden existir materiales topológicos en ausencia de campos magnéticos externos (y que por lo tanto son simétricos bajo IT), en donde el acoplamiento espín-órbita (el acoplamiento relativista entre el momento angular del electrón y su espín) juega un papel fundamental. A estos materiales los llamamos *aislantes topológicos* (AT).

Después del descubrimiento de von Klitzing y la explicación de TKNN sobre el origen topológico del estado HC, surgió la pregunta natural: ¿Puede existir este efecto en ausencia de un campo magnético? El primer paso en esta dirección lo dio Duncan Haldane en 1988, quien concibió un sistema (básicamente electrones moviéndose en una red hexagonal o de panal de abeja) en donde el efecto Hall podía ocurrir en ausencia de un campo magnético externo. Sin embargo, su predicción teórica se anticipó demasiado a las posibilidades experimentales de la época, de manera que su modelo quedó como una mera predicción. El siguiente paso crucial en esta historia lo dieron Kane y Mele en 2005, y Bernevig, Huges y Zhang (BHZ) en 2006, quienes propusieron de forma independiente que dos copias del modelo de Haldane podían dar origen a una nueva fase topológica de la materia, el *estado Hall cuántico de espín* (HCE), que podía realizarse en ciertos modelos teóricos con acoplamiento espín-órbita (EO). El papel del acoplamiento EO en estos modelos puede entenderse como sigue. Clásicamente, si un electrón orbita alrededor del núcleo, en el marco de reposo del electrón, el campo eléctrico generado por el núcleo se siente como un campo magnético, cuya dirección depende del espín del electrón. Para que este campo magnético inducido por el acoplamiento espín-órbita reemplace al campo externo del sistema tipo Hall, es necesario que sea intenso, y para ello, el campo eléctrico también debe ser intenso (sólo presente en núcleos pesados). Esta situación lleva a la aparición de dos estados de borde con helicidad bien definida (las dos proyecciones del espín se propagan en direcciones opuestas), que pueden entenderse como dos copias del efecto HC y son los que dan lugar al efecto HCE. Ver el panel superior de la Fig. 4.

Aunque todos los materiales tiene acoplamientos EO, sólo unos pocos resultan ser aislantes topológicos. BHZ propusieron un mecanismo general para encontrar aislantes topológicos y predijeron que pozos cuánticos de telurito de mercurio son ATs cuando su espesor es mayor que una distancia crítica $d_c$. El mecanismo general que modula los regímenes trivial y topológico es la inversión de bandas, en donde el orden usual de las bandas de valencia y conducción se invierten debido al acoplamiento EO.

En la mayoría de los semiconductores, la banda de conducción esta formada por los orbitales $s$ y la banda de valencia está formada por los orbitales $p$. En algunos elementos pesados, como el mercurio (Hg) y telurio (Te), el acoplamiento EO es tan fuerte que los orbitales $p$ quedan por arriba de los orbitales $s$, esto es, las bandas se invierten. Los pozos cuánticos de telurito de mercurio pueden prepararse intercalando dicho material con telurito de cadmio, que tiene una constante de red similar pero un acoplamiento EO mucho menor. Por lo tanto, incrementando el espesor de la capa de HgTe se incrementa la intensidad del acoplamiento EO en el pozo cuántico completo. Para un pozo cuántico delgado, los efectos del CdTe dominan y las bandas tienen el ordenamiento normal: la sub-banda de conducción tipo $s$, se localiza arriba de la sub-banda de valencia tipo $p$. En un pozo grueso ocurre el orden opuesto debido al incremento del espesor $d$ de la capa de HgTe. El grosor crítico $d_c$ para que ocurra la inversión de las bandas es de alrededor de 6.5nm. En términos prácticos, el aislante topológico bidimensional HgTe está descrito por el Hamiltoniano efectivo

$$H_{HgTe} = \epsilon_0(k) \mathbb{1} + M_k \sigma_z + A(k_x \sigma_x + k_y \sigma_y) \quad (4)$$

donde $\epsilon_k = C - Dk^2$ y $M_k = M - Bk^2$. A, B, C, D y M son parámetros del sistema. El bloque superior de 2 × 2 describe los electrones con espín up en la banda de conducción y en la banda de valencia; mientras que el bloque inferior describe los electrones con espín down en estas bandas. La brecha de energía entre las bandas es $2M$, y $B$, que es típicamente negativo, describe la curvatura de las bandas. Este modelo se puede resolver fácilmente. Para $M/B$ 0, el Hamiltoniano (4) describe un aislante trivial. Pero para pozos cuánticos gruesos, las bandas se invierten, $M$ se hace negativo y la solución describe a los estados del borde del efecto Hall cuántico de espín. La aparición de estos estados metálicos inusuales en el borde es la principal signatura experimental de que un aislante es topológico. A menos de un año de su predicción teórica, el grupo liderado por L. Molenkamp en la Universidad de Würzburg confirmó experimentalmente el efecto HCE, abriendo las puertas así a la computación cuántica de espín.

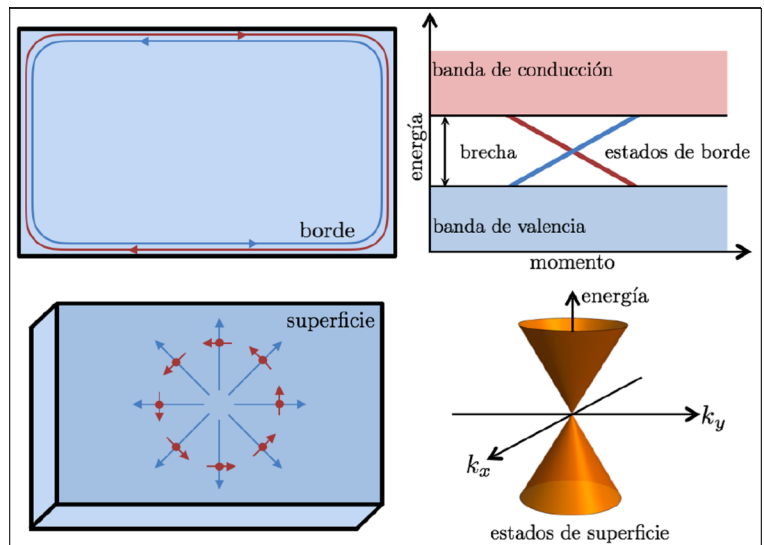

Figura 4: Panel superior: representación esquemática del efecto Hall cuántico de espín y su estructura de bandas. El borde contiene modos de espín opuesto que se propagan hacia la derecha y hacia la izquierda, y éstos se relacionan por la simetría de inversión temporal. Panel inferior: aislante topológico tridimensional y cono de Dirac (relación energía-momento en 2D) de los estados de superficie. La superficie soporta movimiento electrónico en cualquier dirección, pero la dirección del movimiento determina de manera único su espín y viceversa.

El siguiente desarrollo teórico importante, también ocurrido en 2006, fue que a diferencia del estado Hall cuántico de von Klitzing, el estado HCE podía generalizarse a tres dimensiones de una manera sutil. En primera instancia, un AT tridimensional débil puede formarse estratificando versiones bidimensionales del efecto Hall cuántico de espín; sin embargo, el estado resultante no es estable ante el desorden y la física es muy parecida a la del estado en 2D. Existe también el aislante topológico fuerte, que tiene una relación mucho más sutil con el caso en 2D: en dos dimensiones es posible conectar una aislante trivial con uno topológico, rompiendo suavemente la simetría de inversión temporal. Esta interpolación continua también puede usarse para construir una estructura de bandas que respete la simetría de inversión temporal que sea topológicamente no trivial. Este aislante topológico fuerte tiene estados de superficie protegidos y es el que ha sido sujeto de actividad experimental en años recientes. Ver el panel inferior de la Fig. 4. Nuevamente el acoplamiento EO es requerido y debe mezclar las componentes de espín. En otras palabras, no hay manera de crear una AT fuerte en 3D a partir de estados de espín separados, como en el caso 2D. En 2007, Liang Fu y Kane predijeron que la aleación $Bi_{1-x}Sb_x$, para algún rango de composición $x$, debe ser un aislante topológico tridimensional. En 2008, el grupo experimental de la Universidad de Princeton, liderado por Zahid Hasan, utilizó ARPES (angle-resolved photoemision spectroscopy) para observar los estados metálicos en este sistema, confirmando así la existencia de aislantes topológicos tridimensionales.

Al igual que en el caso de los pozos cuánticos de telurito de mercurio, la familia topológicamente no trivial $Bi_2Te_3$ debe su origen a la inversión de bandas, controlada por el acoplamiento EO del $Bi$ y $Te$. Debido a dicha similitud, esta familia puede describirse por una versión 3D del modelo HgTe. El Hamiltoniano correspondiente es:

$$H_{i_2Te_3} \quad 0 \quad (\quad_k \quad 0 \quad M_k\quad_z) \quad A_1K_z\quad_x\quad\quad_x$$
$$A_2(k_x\quad_z\quad\quad_x\quad k_y\quad_0\quad\quad_y), \quad (5)$$

donde

$$_k \quad C \quad D_1k_z^2 \quad D_2k^2 \text{ y } M_k \quad M \quad B_1k_z^2 \quad B_2k^2.$$

Al igual que en el modelo en 2D, la solución para $M/B_1$ 0 describe un aislante trivial, mientras que para $M/B_1$ 0, las bandas se cruzan y el sistema es un AT.

**Teorías topológicas de campos**

En física de materia condensada, a menudo estamos interesados en las propiedades de baja energía (longitud de onda grande) de un sistema. En este régimen, los detalles del Hamiltoniano cuántico no son importantes y las propiedades físicas del sistema pueden describirse en términos de una teoría de campos efectiva. Para estados convencionales clasificados por el rompimiento de simetrías, la teoría efectiva queda determinada por el parámetro de orden, sus simetrías y dimensionalidad. Los estados topológicos de la materia también pueden describirse mediante una teoría de campos efectiva de baja energía. En este caso, la teoría involucra términos topológicos que capturan las propiedades del sistema. Por ejemplo, el efecto HC puede describirse por la teoría de Chern-Simons en 2D, que se define mediante la acción efectiva

# Artículos

$$S_{2D} \sim \frac{n_1}{4} \int d^2x\, dt\, A\, \partial A, \quad (6)$$

donde $n_1$ es el primer número de Chern, $A = (A_0, \mathbf{A})$ es el potencial electromagnético y $\alpha = e^2/\hbar c$ es la constante de estructura fina. Bajo inversión temporal, $A_0 \to A_0$ y $\mathbf{A} \to -\mathbf{A}$, de donde se observa que esta teoría rompe dicha simetría. La respuesta de baja energía del estado Hall cuántico se deriva de esta acción efectiva. Por ejemplo, tomando la derivada funcional de la acción (6) respecto a $A$, obtenemos la corriente $\frac{1}{c} j \sim (n_1 \alpha/2\pi) \partial A$, que describe exactamente al efecto Hall cuántico con conductividad Hall $\sigma_{Hall} \sim n_1(e^2/h)$. De hecho, implica que un campo eléctrico $E$ induce una corriente transversal $J \sim \sigma_{Hall} i \sigma_y E$, y un campo magnético $B$ induce una acumulación de cargas $j^0 \sim (n_1 \alpha/2\pi)B$, donde $\sigma_y$ es la matriz de Pauli.

La teoría efectiva del efecto HC en 2D no sólo captura los aspectos topológicos del sistema, sino que también proporciona una ruta para generalizar el estado HC (que rompe la simetría de IT) a estados topológicos invariantes bajo IT en más dimensiones. El camino hacia la generalización es claro: dado que el término de Chern-Simons puede generalizarse a todas las dimensiones impares, la topología de los estados HC también puede extenderse a esas dimensiones. La generalización del efecto HC a 4D nos da el efecto HC fundamental invariante bajo IT del que se derivan todos los efectos HC de dimensiones menores por el proceso de reducción dimensional. La teoría efectiva que se obtiene en 3D es

$$S \sim \frac{\alpha}{32\pi^2} \int d^3x\, dt\, \theta(x,t) e^{\mu\nu\rho\sigma} F_{\mu\nu} F_{\rho\sigma}, \quad (7)$$

donde $\theta(x,t)$ es un campo de acoplamiento que se deriva del proceso de compactificación. En física de partículas, a esta teoría se le conoce como electrodinámica axiónica y a $\theta(x,t)$ se le llama campo axiónico. Como se discutió, la teoría topológica de Chern-Simons en 4D es invariante bajo IT. Por lo tanto, es natural preguntarse cómo puede preservarse dicha simetría en el proceso de reducción dimensional. Físicamente, $\theta$ representa el flujo magnético del campo de norma en la dimensión extra a través del círculo compactificado, y la física debe ser invariante bajo la traslación $\theta \to \theta + 2\pi$. Además, la simetría de inversión temporal transforma $\theta$ en $-\theta$. Por lo tanto, hay dos valores posibles de $\theta$ que son consistentes con dicha simetría, $\theta = 0$ y $\theta = \pi$. En el último caso, la IT transforma $\theta$ en $-\pi$, que es equivalente a $\pi$ mod $2\pi$. Concluimos así que hay dos clases diferentes de aislantes topológicos invariantes bajo IT en 3D, la clase topológicamente trivial con $\theta = 0$ y la clase topológicamente no trivial con $\theta = \pi$.

## Efecto magnetoeléctrico topológico

Además de sus interesantes propiedades electrónicas, los ATs también exhiben propiedades electromagnéticas de interés que se derivan de la teoría efectiva definida por la Ec. (7). Específicamente, los aislantes topológicos tienen la habilidad de mezclar los campos eléctrico $E$ y de inducción magnética $B$, lo que da lugar a nuevos fenómenos electromagnéticos característicos de dichos materiales. Esta propiedad topológica de los ATs se manifiesta cuando la simetría de IT se preserva en el volumen pero se rompe en la superficie (para abrir una brecha en los estados de superficie). La perturbación puede ser, por ejemplo, una cubierta ferromagnética delgada o un campo magnético externo perpendicular a la superficie. En tal caso, el material se vuelve completamente aislante y el parámetro $\theta$ de la Ec. (7) está cuantizado en múltiplos impares de $\pi$; a saber, $\theta = (2n+1)\pi$, donde $2n+1 \in \mathbb{Z}$ es el número de fermiones de Dirac en la superficie del aislante topológico. Valores positivos o negativos de $\theta$ están relacionados con la dirección de la magnetización (o del campo magnético) en la superficie del material. La respuesta electromagnética de los aislantes topológicos simétricos bajo IT en 3D se describe suplementando la teoría de Maxwell con el término topológico de la Ec. (7). De hecho, las ecuaciones de movimiento que se obtienen son las ecuaciones de Maxwell en materia, con las relaciones constitutivas

$$D = E + (\ /\ )B,$$
$$H = (B/\ ) - (\ /\ )E. \quad (8)$$

Una consecuencia inmediata de las ecuaciones de Maxwell modificadas es que en ausencia de un campo eléctrico, un campo magnético puede generar una polarización eléctrica, y viceversa. A esta habilidad de los aislante topológicos se le conoce como *efecto magnetoeléctrico topológico* (MET), y es una manifestación del orden topológico no trivial de estos materiales. Interesantemente, el efecto MET es el primer fenómeno de cuantización topológica en unidades de la constante de estructura fina. A continuación, discutimos algunas manifestaciones del efecto MET.

### Rotaciones topológicas de la luz

Los efectos Kerr y Faraday son fenómenos magneto-ópticos que resultan de la transferencia de momento angular de una onda que incide sobre un material magnético a las ondas reflejada y transmitida, respectivamente. Este mecanismo resulta en la rotación de los planos de polarización de las ondas reflejada y transmitida con respecto al de la luz incidente, y a los ángulos de rotación se les denomina ángulos de Kerr y Faraday, respectivamente. Dichos efectos magneto-ópticos permiten sondear directamente el rompimiento de la simetría de inversión temporal en sólidos.

Como se mencionó antes, cuando la simetría IT se rompe débilmente en un AT, se abre una brecha en el espectro de energía de los estados de superficie del material y éste exhibe el efecto magnetoeléctrico topológico. La mezcla entre los campos electromagnéticos debido al término axiónico da origen a efectos Kerr y Faraday en este tipo de materiales; sin embargo, estos no resultan de propiedades magnéticas específicas, sino de la topología no trivial de la estructura de bandas de dichos materiales. Es decir, estos efectos son una firma única del efecto magnétoeléctrico topológico de ATs en 3D.

### Monopolos magnéticos

Las ecuaciones que rigen todos los fenómenos electromagnéticos fueron formuladas por el físico escocés James Clerk Maxwell en 1864. La ley de Gauss, $E = 4$ , expresa que cargas eléctricas puntuales son fuente de campo eléctrico. Sin embargo, la ecuación $B = 0$ refleja que no existen fuentes monopolares de campo magnético. En 1931, Paul A. M. Dirac demostró que la existencia de monopolos magnéticos es consistente con la teoría cuántica. A pesar de éste y muchos otros esfuerzos concernientes a la existencia de los monopolos magnéticos, aún no se ha detectado esta elusiva partícula, pero de acuerdo a Dirac, si podría existir, tan sólo que quizás no la hemos encontrado. Nacen entonces preguntas fundamentaales que han inquietado a la comunidad científica: ¿existen los monopolos magnéticos? y si es así ¿dónde están que no los vemos con la misma facilidad que vemos, por ejemplo, electrones?

Recientemente hubo revuelo por la observación de excitaciones tipo monopolo en un tipo de materiales conocidos como hielos de espín, aunque existe cierto escepticismo con el resultado. En los hielos de espín, los grados de libertad originales (dipolos magnéticos) se fraccionalizan en polos magnéticos aislados atados por pares. El reciente descubrimiento de los ATs provee un esquema físico distinto, en donde la existencia de campos magnéticos monopolares está permitida. Consideremos, por ejemplo, una carga eléctrica puntual cerca de la superficie de un aislante trivial. La carga eléctrica polarizará el dieléctrico, que se puede describir mediante la aparición de una carga eléctrica imagen al interior del material. Si hacemos lo mismo con un aislante topológico, cuyos estados de superficie presentan una brecha inducida por una perturbación que rompa la simetría de inversión temporal, además de la carga eléctrica imagen también aparecerá un monopolo magnético imagen al interior del material. La aparición de dichos monopolos no viola las leyes de Maxwell, pues se trata de una excitación emergente que describe los efectos físicos asociados a la corriente de Hall que se induce en la superficie de los ATs, y no de la partícula fundamental en el sentido de Dirac. Esta propiedad exótica de los ATs podría usarse para escribir memorias magnéticas mediante medios puramente eléctricos.